\documentclass[journal,12pt,onecolumn,draftclsnofoot,]{IEEEtran}

\usepackage[T1]{fontenc}

\usepackage{amsmath}
\usepackage{cite}

\usepackage{graphicx}
\usepackage{url}

\usepackage{epstopdf}
\usepackage{amsmath}
\usepackage{amsxtra}
\usepackage{amstext}
\usepackage{amssymb}
\usepackage{latexsym}
\usepackage{dsfont} 
\usepackage{color}

\usepackage{multirow}
	
\usepackage{float}

\usepackage{hyperref}

\usepackage{units}

\usepackage{cases}

\newcommand{\mbf}[1]{\mathbf{#1}}

\newcommand{\nth}[1]{{#1}{\text{th}}}

\newcommand{\abs}[1]{\left|{#1}\right|}
\newcommand{\norm}[1]{\left\|{#1}\right\|}

\newcommand\Tstrut{\rule{0pt}{2.3ex}}         

\DeclareMathOperator*{\argmax}{arg\,max}   
\DeclareMathOperator*{\argmin}{arg\,min}

\newcommand{\ML}{\mathrm{ML}}

\newcommand{\ZF}{\mathrm{ZF}}

\usepackage{algorithm}
\usepackage{algpseudocode}

\makeatletter
\def\BState{\State\hskip-\ALG@thistlm}
\makeatother

\begin{document}

\title{Modulation Classification via Subspace Detection in MIMO Systems}

\author{Hadi~Sarieddeen,~\IEEEmembership{Student Member,~IEEE,}
        Mohammad~M.~Mansour,~\IEEEmembership{Senior Member,~IEEE,}
        and~Ali~Chehab

}

\maketitle

\begin{abstract}
The problem of efficient modulation classification (MC) in multiple-input multiple-output (MIMO) systems is considered. Per-layer likelihood-based MC is proposed by employing subspace decomposition to partially decouple the transmitted streams. When detecting the modulation type of the stream of interest, a dense constellation is assumed on all remaining streams. The proposed classifier outperforms existing MC schemes at a lower complexity cost, and can be efficiently implemented in the context of joint MC and subspace data detection.
\end{abstract}


\IEEEpeerreviewmaketitle

\section{Introduction}

\IEEEPARstart{M}{odulation} classification is the task of recognizing the modulation type (MT) employed at the transmitter of a detected signal, which is required for various military and civilian applications. In particular, cognitive radio with adaptive MTs \cite{arslan2007cognitive} is a promising future application of MC. In such scenario, the transmitter dynamically adjusts the data rate by switching the modulation order depending on channel conditions. By employing automatic/blind MC at the receiver, the communication overhead can be reduced.

MC techniques are of two types: feature-based and likelihood-based \cite{Dobre}. While adding more antennas in MIMO technology \cite{2003_Paulraj} increases spectral efficiency, the complexity of optimal likelihood-based MC \cite{Eldemerdash2016,Choqueuse,Sarieddeen_ICASSP_MC} increases exponentially with this addition, especially when each layer is allowed to have a different MT. An alternative low complexity sub-optimal feature-based MC scheme for MIMO systems exploits the higher-order cyclic cumulants (CCs) of the baseband intercepted signal \cite{Muhlhaus2013}. Moreover, without perfect channel state information (CSI), independent component analysis has been used \cite{Muhlhaus2013b} to blindly estimate the channel in conjunction with either likelihood-based or feature-based MC.

In this letter, we propose near-optimal low-complexity likelihood-based MC for MIMO systems. We first decompose the channel matrix employing subspace decomposition, and then detect the MT on the partially decoupled stream of interest using a modified likelihood metric. A joint MC and subspace data detection receiver is also proposed. Regarding notations, bold upper case, bold lower case, and lower case letters correspond to matrices, vectors, and scalars, respectively. Scalar norm, vector norm, conjugate transpose, and inverse are represented by $\abs{\cdot}$, $\norm{\cdot}$, $(\cdot)^{*}$, and $(\cdot)^{-1}$, respectively.

\section{System Model}
\label{sec:sysmodel}

We consider, a spatial multiplexing MIMO system with $N$ transmit, and for simplicity $M\!=\!N$ receive antennas:
\begin{equation}\label{eq:sysmodel}
  \mbf{y} = \mbf{H}\mbf{x} + \mbf{z}
\end{equation}
with $\mbf{y}\!=\![y_{1}y_{2}\ldots y_{M}]^{T}\!\in\!\mathcal{C}^{M\!\times\!1}$ being the received complex vector, $\mbf{H}\!=\![\mbf{h}_{1}\mbf{h}_{2}\ldots \mbf{h}_{N}]\!\in\!\mathcal{C}^{M\!\times\!N}$ the complex channel matrix, $\mbf{x}\!=\![x_{1}x_{2}\ldots x_{N}]^{T}\!\in\!\mathcal{C}^{N\!\times\!1}$ the transmitted symbol vector, and $\mbf{z}\!\in\!\mathcal{C}^{M\!\times\!1}$ the complex additive white Gaussian noise vector with zero mean and variance $\sigma^{2}$ \big($\mathsf{E}[\mbf{z}\mbf{z}^{*}]=\sigma^{2}\mbf{I}_N$\big). Each symbol $x_{n}$ belongs to a normalized complex constellation $\mathcal{X}_{n}$ of size $Q_{n}\!=\!2^{q_{n}}$, thus $\mbf{x}\!\in\! \mathcal{\bar{X}}\!=\!\mathcal{X}_{1}\times \ldots \times\mathcal{X}_{N}$ and $\mathsf{E}[x_{n}^{*}x_{n}]\!=\!1$. Consequently, the signal to noise ratio ($\mathsf{SNR}$) is defined in terms of the noise variance as $\mathsf{SNR}\!=\!(N/\sigma^{2})$. The bit representation of symbol $x_{n}$ is a coded bit-interleaved sequence $\mbf{b}_{n}\!=\!(b_{n,1},b_{n,2},\ldots,b_{n,q_{n}})$.

In a MIMO system that supports non-uniform MTs, each of the $N$ transmitted symbols is assumed to be drawn from one of $S$ possible MTs, with equal probability. We develop MC schemes to estimate the MT per-layer, using the received signal $\mbf{y}$ and assuming perfect CSI.

\section{Likelihood-Based MC}
\label{sec:likelihoodbased}

The optimal likelihood-based MC scheme decides on the MT that has the maximum likelihood within multiple hypotheses. Bayesian hypothesis testing is performed on the $S^N$ possible hypotheses, corresponding to $\mathcal{\bar{X}}_j\!=\!\mathcal{X}_{j,1}\times \ldots \times\mathcal{X}_{j,N}$ finite lattices ($j\in\{1,\ldots,S^N\}$), with likelihoods:
\begin{equation}\label{eq:gen}
P(\mbf{y};\mathcal{\bar{X}}_j)=\sum_{\mbf{x}\in\mathcal{\bar{X}}_j} P(\mbf{y}|\mbf{x})P(\mbf{x})
\end{equation}
where $P$ is the probability density function. Under statistical independence between the components of $\mbf{x}$, and assuming uniform priors, $P(x_{n})\!=\!1/\abs{\mathcal{X}_n}$, where $\abs{\cdot}$ denotes the constellation set cardinality, the decision metric is derived as:
\begin{equation}\label{eq:gen2}
\eta = \argmax_{j\in\{1,\ldots,S^N\}}\sum_{\mbf{x}\in\mathcal{\bar{X}}_j} P(\mbf{y}|\mbf{x})\frac{1}{\abs{\mathcal{X}_{j,1}}}\times\cdots\times\frac{1}{\abs{\mathcal{X}_{j,N}}}
\end{equation}

Noting that $P(\mbf{y}|\mbf{x})\!=\!\frac{1}{({\pi\sigma^{2}})^{M}} \exp(-\frac{1}{\sigma^{2}} \norm{\mbf{y} - \mbf{Hx}}^{2})$, and neglecting the term $\frac{1}{({\pi\sigma^{2}})^{M}}$ which is assumed fixed over hypotheses, the resultant Log-MAP decision metric is:
\begin{equation}\label{eq:logmap}
\eta_{\text{L}} = \argmax_{j\in\{1,\ldots,S^N\}} \bigg(\log\frac{1}{\abs{\mathcal{X}_{j,1}}} + \cdots + \log\frac{1}{\abs{\mathcal{X}_{j,N}}} + \log\sum_{\mbf{x}\in\mathcal{\bar{X}}_j}  \exp\big(-\frac{1}{\sigma^{2}} \norm{\mbf{y} - \mbf{Hx}}^{2}\big) \bigg)
\end{equation}
which is the average likelihood ratio test (ALRT) solution.

Solving equation \eqref{eq:logmap} is computationally intensive, because for each $j$ we have to calculate $\abs{\mathcal{X}_{j,1}}\times\cdots\times\abs{\mathcal{X}_{j,N}}$ exponential terms. However, one of these terms is dominant and corresponds to the scaled maximum likelihood (ML) distance:
\begin{equation}
d_{\ML,j} = \min_{\mbf{x}\in\mathcal{\bar{X}}_j} \frac{1}{\sigma^{2}} \norm{\mbf{y} - \mbf{Hx}}^{2}
\end{equation}
Using the approximation $\log\sum_{r}{\exp(a_{r})}\approx\max_{r}\{a_{r}\}$, we get:
\begin{equation}\label{eq:maxlogmap}
\eta_{\text{M}} = \argmax_{j\in\{1,\ldots,S^N\}} \bigg(\log\frac{1}{\abs{\mathcal{X}_{j,1}}} + \cdots + \log\frac{1}{\abs{\mathcal{X}_{j,N}}} - d_{\ML,j} \bigg)
\end{equation}
which is the near-optimal Max-Log-MAP classifier.

While Max-Log-MAP eliminates exponential operations, the number of Euclidean distance computations per hypothesis remains exponential, and computing the likelihood functions $S^N$ times is exhaustive. An alternative approach is required, that separates the transmitted signals for individual treatment, which results in only $\abs{\mathcal{X}_{j,n}}$ distance computations per layer $n$ and hypothesis $j\in\{1,\ldots,S\}$. This is achieved by the per-layer sub-optimal ALRT solution. With perfect CSI at the receiver, the sub-optimal ALRT classifier finds the zero-forcing (ZF) equalized output $\mbf{\hat{y}}_{\ZF} = \big(\mbf{H}^{*}\mbf{H})^{-1} \mbf{H}^{*}\mbf{y}$, computes the scaled noise variance $\sigma^{2}_{\ZF}\!=\!(\mbf{h}_n^{*}\mbf{h}_n)^{-1}\sigma^{2}$, and generates the likelihood function per layer $n$ as follows:

\begin{equation}\label{eq:zflogmap}
\eta_{\text{S}} =\argmax_{j\in\{1,\ldots,S\}} \bigg(\log\frac{1}{\abs{\mathcal{X}_{j,n}}} + \log\sum_{x_n\in\mathcal{X}_{j,n}}  \exp\big(-\frac{1}{\sigma^{2}_{\ZF}} \abs{\hat{y}_{ZF,n} - x_n}^{2}\big) \bigg)
\end{equation}

We seek a classifier that decouples the layers while maintaining distance metrics that are close to that of Log-MAP.

\section{Proposed MC Scheme}
\label{sec:proposed}

\begin{figure}[t]
\centering
\includegraphics[width=4in]{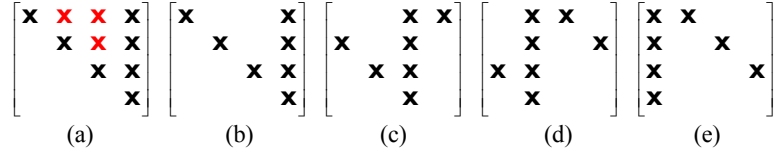}\vspace{-0.1in}
\caption{4x4 channel matrix structures}
\label{f:matrices}
\end{figure}

\subsection{Subspace Decomposition}
\label{sec:subspace}

While the standard QR decomposition (QRD) decomposes the channel matrix $\mbf{H}$ into an upper triangular matrix (UTM) $\mbf{R}=[u_{pq}]\in\mathcal{C}^{N\times N}$ with $u_{pp}\in\mathcal{R}^{+}$, and a unitary matrix $\mbf{Q}$, the so-called WR decomposition (WRD) scheme punctures in addition the red-marked entries above the diagonal in Fig. \ref{f:matrices}(a). We aim at transforming $\mbf{H}$ into a punctured UTM $\mbf{R}$ as shown in Fig. \ref{f:matrices}(b), through a matrix $\mbf{W}=[\mbf{w}_{1}\mbf{w}_{2} \ldots \mbf{w}_{N}]\in\mathcal{C}^{N\times N}$, such that $\mbf{W}^{*}\mbf{H}=\mbf{R}$. We assume $\mbf{H}=[\mbf{h}_{1}\mbf{h}_{2}\ldots\mbf{h}_{N}]$ to have a full column rank, and impose the condition on the column vectors of $\mbf{W}$ to have unit length, i.e., $\mbf{w}_{n}^{*}\mbf{w}_{n}\!=\!1$ for $n\!=\!1,\ldots,N$, so that the transformed noise vector will maintain an unaltered covariance matrix ($\mathsf{E}[\mbf{W}^{*}\mbf{zz}^{*}\mbf{W}]=\sigma^{2}\mbf{I}_{N}$).

The brute force approach for computing $\mbf{W}$ involves extensive matrix inversions, which is computationally intensive and prone to numerical error. An alternative approach \cite{2014_mansour_SPL_WLD} consists of QRD followed by elementary matrix operations. Let $\mbf{H}$ be QR-decomposed such that $\mbf{Q}_1^{*}\mbf{H}=\mbf{R}_{1}$. Consider row $1\!<n\!\leq N$ of $\mbf{R_{1}}$, and assume the $\nth{m}$ entry $r_{nm}$, $m\!>\!n$, is to be nulled. We have $\mbf{q}_{n}^{*}\mbf{h}_{m}=r_{nm}\in\mathcal{C}$ and $\mbf{q}_{m}^{*}\mbf{h}_{m}=r_{mm}\in\mathcal{R}^{+}$, from which it follows that $(\mbf{q}^{*}_{n}-\mbf{q}^{*}_{m}\frac{r_{nm}}{r_{mm}})\mbf{h}_{m}=0$. Therefore:
\begin{equation}\label{eq:punc1}
  \mbf{q}_{n}=\mbf{q}_{n}-\mbf{q}_{m}r^{*}_{nm}/r_{mm}
\end{equation}
\begin{equation}\label{eq:punc2}
  r_{nj}=r_{nj}-r_{mj}r_{nm}/r_{mm},\ \ \text{for}\ j=m,\ldots,N
  \end{equation}
puncture the required entry and update $\mbf{Q}_{1}$ accordingly. Finally, the non-zero entries in row $n$ of $\mbf{R}_{1}$ are updated, and $\mbf{q}_{n}$ is normalized to have unit length:
\begin{equation}\label{eq:punc3}
  r_{nj}=r_{nj}/\norm{\mbf{q}_{n}}, \ \ \text{for}\ j=n,\ldots,N
\end{equation}
\begin{equation}\label{eq:punc4}
  \mbf{q}_{n}=\mbf{q}_{n}/\norm{\mbf{q}_{n}}
\end{equation}
Repeating this on required entries (bottom to top, right to left), the resulting $\mbf{Q}_{1}$ is $\mbf{W}$, and $\mbf{R}_{1}$ is the desired punctured $\mbf{R}$.

\subsection{Proposed Likelihood-Based MC}
\label{sec:proposed_MC}

To generate the likelihood functions on all layers, the $N$ streams are decoupled, one at a time, by cyclically shifting the columns of $\mbf{H}$ and generating the punctured UTMs, as shown in Fig. \ref{f:matrices}(b-e). Alternatively, a minimal swapping operation can put the layer of interest at the rightmost column location, and hence a decomposition as shown in Fig. \ref{f:matrices}(b) will always follow. We represent this swapping operation by a permutation:
\begin{equation}\label{eq:permutation}
\pi^{(n)}(i)=\left\{
                \begin{array}{ll}
                  N \ \ \text{if} \  i=n\\
                  n \ \ \text{if} \  i=N\\
                  i \ \ \text{otherwise}
                \end{array}
              \right.
\end{equation}
for $i\!=\!1,\ldots,N$ and at a layer of interest $n$. Each permuted $\mbf{H}^{(n)}$ is then WR-decomposed into $\mbf{W}^{(n)}$ and $\mbf{R}^{(n)}$.

We first partition $\mbf{\tilde{y}}^{(n)}$, $\mbf{R}^{(n)}$, and permuted $\mbf{x}$ as:
\begin{equation}\label{eq:partition}
\mbf{\tilde{y}}^{(n)}=\begin{bmatrix} \mbf{\tilde{y}}^{(n)}_{1} \\ \tilde{y}^{(n)}_{2}  \end{bmatrix}, \ \ \mbf{R}^{(n)}=\begin{bmatrix} \mbf{A}^{(n)} & \mbf{b}^{(n)} \\ 0 & c^{(n)} \end{bmatrix}, \ \  \mbf{x}=\begin{bmatrix} \mbf{x}_{1} \\ \tilde{x}_{2}  \end{bmatrix}
\end{equation}
where $\mbf{\tilde{y}}^{(n)}_{1}\!\in\!\mathcal{C}^{(N-1)\times1}$, $\tilde{y}^{(n)}_{2}\!\in\!\mathcal{C}^{1\times1}$, $\mbf{A}^{(n)}\!\in\!\mathcal{R}^{(N-1)\times(N-1)}$, $\mbf{b}^{(n)}\!\in\!\mathcal{C}^{(N-1)\times1}$, $c^{(n)}\!\in\!\mathcal{R}^{1\times1}$, $\mbf{x}_{1}\!\in\!\mathcal{X}^{N-1}$, and $\tilde{x}_{2}\!\in\!\mathcal{X}_n$.
Then, the modified distance metric is expressed in terms of $\tilde{x}_{2}$ as:

\begin{equation}\label{eq:dist}
\norm{\mbf{\tilde{y}}^{(n)}\!-\!\mbf{R}^{(n)}\mbf{x}}^{2}\!=\!\abs{\tilde{y}_{2}^{(n)}\!-\!c^{(n)}\tilde{x}_{2}}^{2}\!+\!\norm{\mbf{\tilde{y}}_{1}^{(n)}\!-\!\mbf{A}^{(n)}\mbf{\hat{x}}_{1}\!-\!\mbf{b}^{(n)}\tilde{x}_{2}}^{2}
\end{equation}

\begin{equation}\label{eq:slicer}
\mbf{\hat{x}}_{1}=\lfloor(\mbf{\tilde{y}}_{1}^{(n)}\!-\!\mbf{b}^{(n)}\tilde{x}_{2})/\mbf{A}^{(n)}\rceil_{\mathcal{X}^{N-1}}
\end{equation}
where $\lfloor \alpha \rceil_{\mathcal{X}_r} \triangleq \argmin_{x_r \in \mathcal{X}_r} \abs{\alpha-x_r}$ is the slicing operator. Since $\mbf{A}^{(n)}$ is a diagonal matrix, slicing is applied to the individual elements of the vector $\mbf{\tilde{y}}^{(n)}_{1}$ in parallel.

Accumulating $T$ observations before deciding on a winning hypothesis, the proposed likelihood functions at layer $n$ are:

\begin{equation}\label{eq:sublogmap}
\hat{\eta}_{\text{L}}=\argmax_{j\in\{1,\ldots,S\}} \sum_{t=1}^{T} \bigg(\log\frac{1}{\abs{\mathcal{X}_{j,n}}} + \log\sum_{\tilde{x}_2\in\mathcal{X}_{j,n}} \exp\big(-\frac{1}{\sigma^{2}} \norm{\mbf{\tilde{y}}^{(n)}\!-\!\mbf{R}^{(n)}\mbf{x}}^{2}\big) \bigg)
\end{equation}

\begin{equation}\label{eq:submaxlogmap}
\hat{\eta}_{\text{M}}=\argmax_{j\in\{1,\ldots,S\}} \sum_{t=1}^{T} \bigg(\log\frac{1}{\abs{\mathcal{X}_{j,n}}} -  \hat{d}_{\ML,j} \bigg)
\end{equation}

\begin{equation}
\hat{d}_{\ML,j} = \min_{\tilde{x}_2\in\mathcal{X}_{j,n}} \frac{1}{\sigma^{2}} \norm{\mbf{\tilde{y}}^{(n)}\!-\!\mbf{R}^{(n)}\mbf{x}}^{2}
\end{equation}

Note that optimal slicing in equation \eqref{eq:slicer} requires knowledge of MTs on all remaining layers, which is infeasible in an independent per-layer scheme. Therefore, we propose to do slicing assuming dense constellations, $\mathcal{X}\!=$1024-QAM for example. The idea of slicing over a dense constellation comes from the work in \cite{Sarieddeen_ICASSP_MC} on multiuser MIMO (MU-MIMO) detection, where it has been shown that near-optimal data detection can be achieved while assuming interferers to have high order MTs, which captures the geometry of constellations while minimizing errors. Note that 1024-QAM was not considered in \cite{Sarieddeen_ICASSP_MC} because of the entailed complexity of ML detection. This is not an issue in our case, since subspace detection only employs the dense constellations in slicing operations.

Table \ref{table:complexity} compares the upper bound on computational complexities of studied classifiers, in terms of the number of Euclidean distance computations, as well as exponential and logarithmic operations, where $\mathcal{X}_{\text{max}}$ is the largest possible MT. Note that the table does not account for the less significant preprocessing computations (ZF equalization, QRD/WRD) that can be computed once for a large number of observations when the channel variation is slow. While computations in optimal ALRT are exponential in the number of transmit antennas, they are linear in the proposed subspace-based classifiers and sub-optimal ALRT solution (the latter is less complex since its distance computations are one dimensional).

\begin{table}[!t]
\caption{Computational complexities of MC schemes}
\label{table:complexity}
\centering
\begin{tabular}{| c || c | c | c | }
\hline
Approach & Euc. Dist. & Exp. & Log. \\
\hline

Log-MAP (ALRT) & $S^N\times\abs{\mathcal{X_{\text{max}}}}^N$ & $S^N\times\abs{\mathcal{X_{\text{max}}}}^N$ & $S^N$ \Tstrut \\ %

Max-Log-MAP & $S^N\times\abs{\mathcal{X_{\text{max}}}}^N$ & $0$ & $0$ \Tstrut \\ %

Sub-optimal ALRT & $N\times S\times\abs{\mathcal{X_{\text{max}}}}$ & $N\times S\times\abs{\mathcal{X_{\text{max}}}}$ & $S$ \Tstrut \\ %

Subspace-Log-MAP & $N\times S\times\abs{\mathcal{X_{\text{max}}}}$ & $N\times S\times\abs{\mathcal{X_{\text{max}}}}$ & $S$ \Tstrut \\

Subspace-Max-Log-MAP & $N\times S\times\abs{\mathcal{X_{\text{max}}}}$ & $0$ & $0$ \Tstrut \\
\hline
\end{tabular}
\end{table}

\section{Joint MC and Subspace Detection}
\label{sec:joint}

In order to generate the log-likelihood ratio (LLR) of the $\nth{k}$ bit of the $\nth{n}$ symbol $b_{n,k}$ via subspace detection, we compute two distance metrics defined as:
\begin{equation}\label{eq:newLLR1}
  \mbf{u}_{n,k}\!=\!\argmin_{x_n \in \mathcal{X}_{n,k}^{0}}{\norm{\mbf{\tilde{y}}^{(n)}\!-\!\mbf{R}^{(n)}\mbf{x}}^{2}},  \mbf{v}_{n,k}\!=\!\argmin_{x_n \in \mathcal{X}_{n,k}^{1}}{\norm{\mbf{\tilde{y}}^{(n)}\!-\!\mbf{R}^{(n)}\mbf{x}}^{2}}
\end{equation}
where $k\!=\!1,\ldots,q_{n}$, the sets $\mathcal{X}_{n,k}^{(0)}\!=\!x_n \in \mathcal{X}_n: b_{n,k}\!=\!0$ and $\mathcal{X}_{n,k}^{(1)}\!=\!x_n \in \mathcal{X}_n: b_{n,k}\!=\!1$ correspond to subsets of symbol vectors in $\mathcal{X}_n$, having in the corresponding $\nth{k}$ bit of the $\nth{n}$ symbol a value of $0$ and $1$, respectively, and the distance metrics are expanded as in equation \eqref{eq:dist}. The unscaled LLRs are then calculated as:
\begin{equation}\label{eq:LLRfinal}
  \Lambda_{n,k} = \norm{\mbf{\tilde{y}}^{(n)} - \mbf{R}^{(n)}\mbf{u}_{n,k}}^{2} - \norm{\mbf{\tilde{y}}^{(n)} - \mbf{R}^{(n)}\mbf{v}_{n,k}}^{2}
\end{equation}

If $\mbf{R}$ is unpunctured, the parallel slicing operation of equation \eqref{eq:slicer} can not be executed, and successive interference cancellation (SIC) is applied to obtain $\mbf{\hat{x}}_{1}$. Expanding the distances in equation \eqref{eq:LLRfinal} via SIC results in a special subspace detector called layered orthogonal lattice detector (LORD) \cite{Siti-1}, and expanding the distances in equations \eqref{eq:sublogmap} and \eqref{eq:submaxlogmap} accordingly results in LORD-Log-MAP and LORD-Max-Log-MAP classifiers, respectively.

Be it detection or MC, while independently processing a layer of interest, the MTs on the remaining layers are unknown, and parallel slicing or SIC is conducted assuming 1024-QAM. This means that the distance metrics computed for data detection are identical to those computed in equations \eqref{eq:sublogmap} and \eqref{eq:submaxlogmap} for the winning hypothesis, and thus combining MC and detection results in a minimal MC overhead.

\begin{figure}[t]
\centering
\includegraphics[width=3.5in]{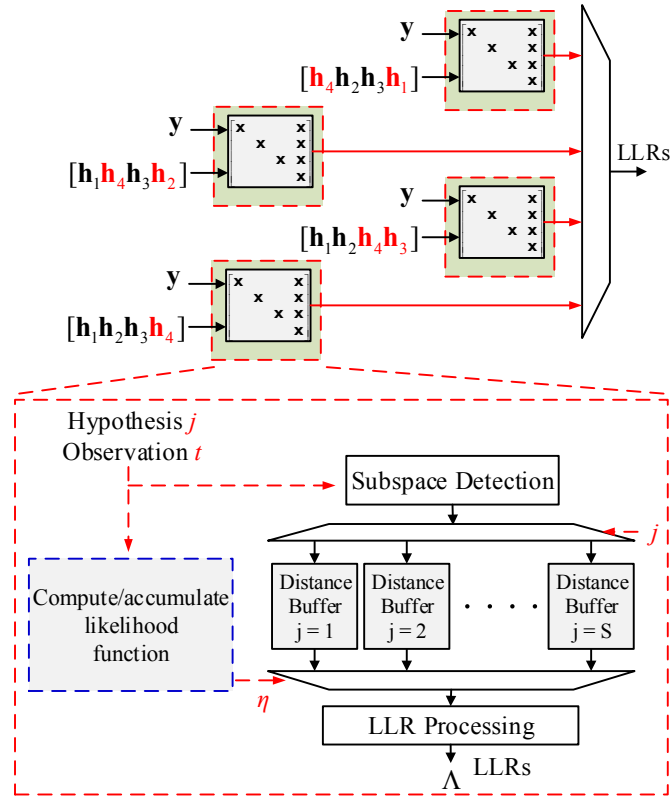}\vspace{-0.1in}
\caption{Joint MC and subspace detection architecture}
\label{f:architecture}
\end{figure}

The joint MC and detection setup is summarized in algorithm \ref{propalg} and architecturally illustrated in Fig. \ref{f:architecture}. For $T$ observations, the detection routine is called $T$ times for all hypotheses, and the resulting distance metrics are stored in memory. Concurrently, the likelihood of each hypothesis is computed. Eventually, the metrics corresponding to the winning hypothesis are retrieved for LLR processing. The receiver can run in this joint mode for a sufficient number of observations and then switch back to regular data detection. Moreover, since the operations on different layers are independent, the proposed algorithm can be parallelized on multiple processing units. Finally, if the system had more receive antennas ($M\!>\!N$), the ``thin'' form of the QR decomposition for tall matrices can be used, and other modifications immediately follow.

\begin{algorithm}[t]
\caption{Proposed per-layer joint MC and detection}\label{propalg}
\begin{algorithmic}[1]
\State Swap the column of interest $n$ with column $N$ in $\mbf{H}$ as in equation \eqref{eq:permutation}
\State Decompose the channel matrix as in equation \eqref{eq:partition}
\State Calculate the distance metrics for all hypotheses as in equation \eqref{eq:dist} while assuming the MTs on the remaining layers to be 1024-QAM
\State Calculate the classifier likelihood function as in equations \eqref{eq:sublogmap} and \eqref{eq:submaxlogmap}
\State Repeat steps 3 and 4 for $T$ observations, accumulate likelihoods, and decide on the winning hypothesis
\State Forward the distance metrics that correspond to the winning hypothesis for bit LLR generation as in equation \eqref{eq:LLRfinal}
\end{algorithmic}
\end{algorithm}

\section{Simulation Results}
\label{sec:simulation}

Several MC and detection schemes were simulated in the context of $4\!\times\!4$ MIMO. We considered five hypotheses of MTs per layer, varying with equal probability on every new frame, which are ${\phi}$, QPSK, 16-QAM, 64-QAM and 256-QAM, with ${\phi}$ representing a constellation having one entry of zero power, corresponding to the case when the transmitting antenna is silent. Note that an all-QAM set of hypotheses that only differs by modulation order is hard to classify, but is more likely to occur in future standards. The winning hypothesis was decided after accumulating $T\!=\!1000$ observations. Turbo coding was used, with a code rate of $1/2$ and $8$ decoding iterations. Moreover, in addition to the regular channel $\mbf{H}$, we considered a correlated channel $\mbf{H}_{c}\!=\!\mbf{R}_{r}^{1/2}\mbf{H}\mbf{R}_{t}^{1/2}$, where $\mbf{R}_{t}$ and $\mbf{R}_{r}$ are the transmit and receive antenna correlation matrices, respectively, with a correlation factor of 0.3.

\begin{figure}[t]
\centering
\includegraphics[width=4.5in]{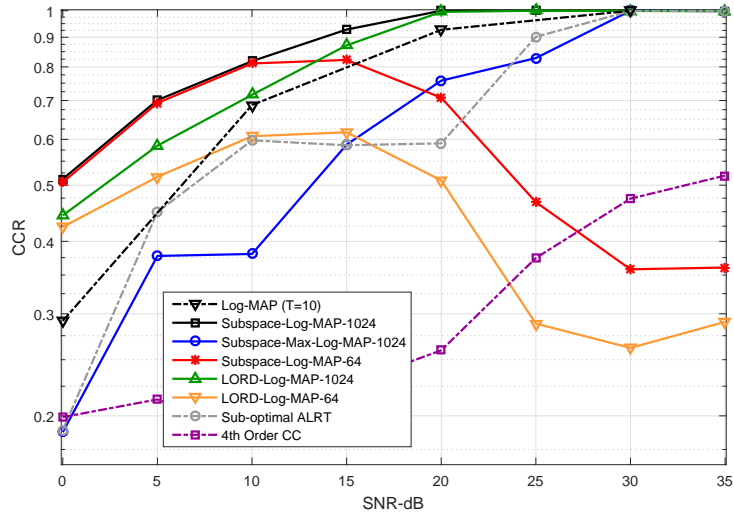}\vspace{-0.1in}
\caption{CCR performance - uncorrelated channels}
\label{f:plot1}
\end{figure}

\begin{figure}[t]
\centering
\includegraphics[width=4.5in]{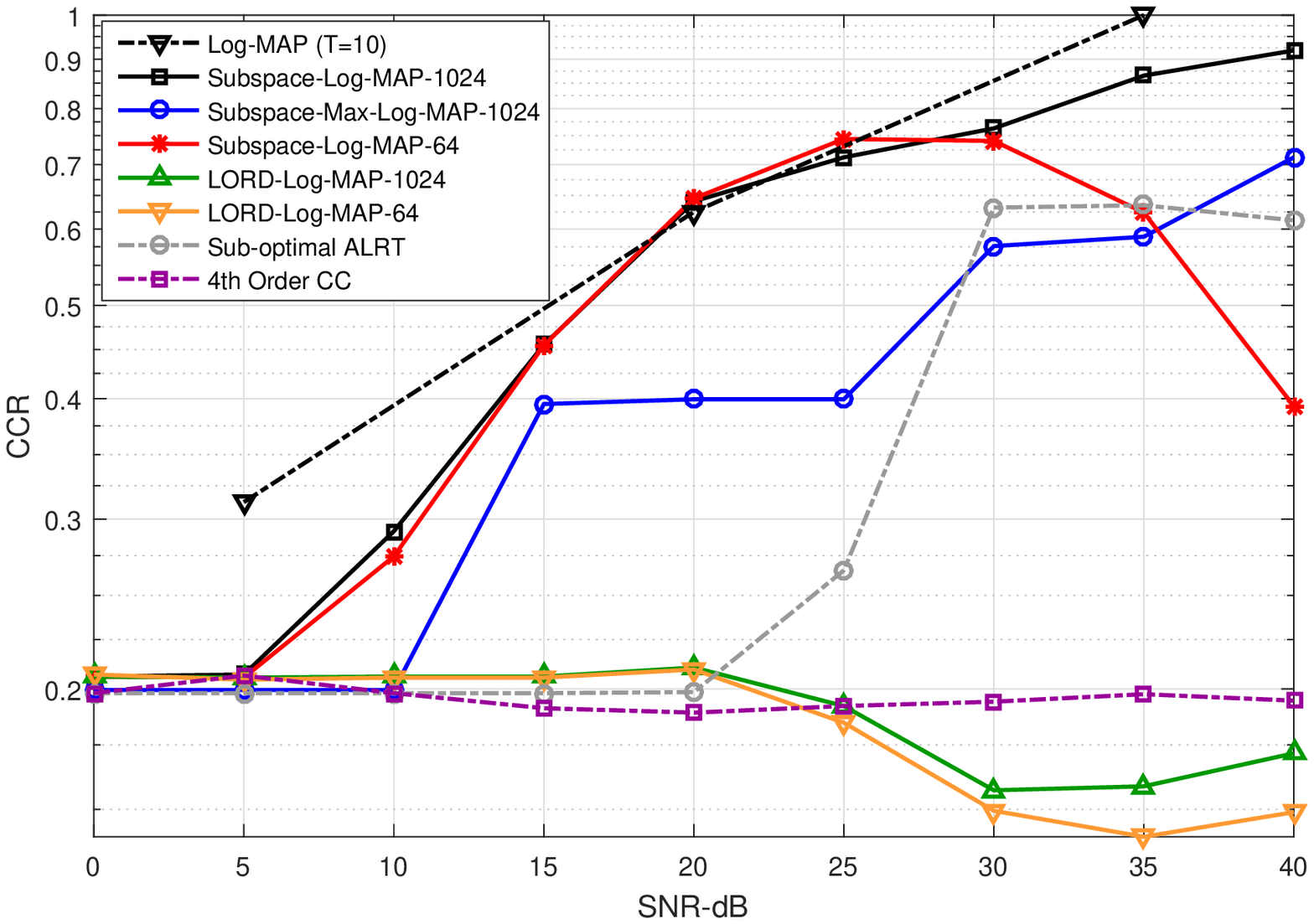}\vspace{-0.1in}
\caption{CCR performance - correlated channels}
\label{f:plot2}
\end{figure}

\begin{figure}[t]
\centering
\includegraphics[width=4.5in]{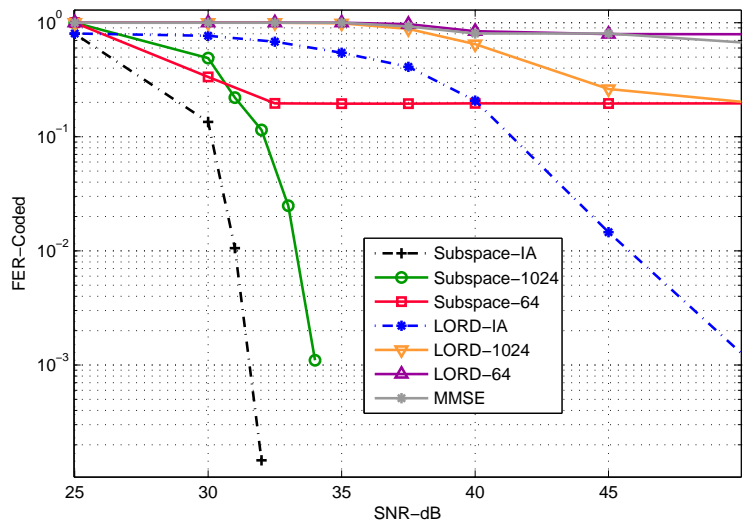}\vspace{-0.1in}
\caption{FER performance - correlated channels - 64-QAM on layer of interest}
\label{f:plot3}
\end{figure}

The proposed classifiers are compared in terms of correct classification ratio (CCR). Fig. ~\ref{f:plot1} shows that for uncorrelated channels, the best performance is achieved by Subspace-Log-MAP and LORD-Log-MAP classifiers, when the MTs on the remaining layers are assumed to be 1024-QAM. The Subspace-Max-Log-MAP and sub-optimal ALRT classifiers lag behind, but are also capable of achieving unity CCR at high $\mathsf{SNR}$. However, assuming 64-QAMs instead of 1024-QAMs resulted in bad classification performance for both subspace and LORD based classifiers. The exhaustive Log-MAP classifier with $T=10$ observations only was added as a reference, and the much less complex proposed approaches outperformed it. Also, the reference feature based (\nth{4} order CCs) classifier performed very bad with an all-QAM set of hypotheses. Fig. ~\ref{f:plot2} then shows that for highly correlated channels, only Subspace-Log-MAP performs well at high $\mathsf{SNR}$, approaching the upper Log-MAP bound.

The corresponding coded frame error rate (FER) performance of the proposed detectors with high channel correlation is shown in Fig. ~\ref{f:plot3}. The detectors were simulated assuming the layer of interest to use 64-QAM (following successful per-layer MC), while the MTs at the remaining layers were unknown, and randomly hopping over possible hypotheses. Both LORD and subspace detectors were tested, assuming the remaining MTs to be 1024-QAM or 64-QAM. These detectors were compared to the regular MT-aware LORD and subspace detectors that have perfect knowledge of MTs on all layers. While regular subspace detection beats LORD by more than $\unit[10]{dB}$, only assuming 1024-QAM in conjunction with subspace detection was able to achieve near MT-aware performance. This declares the subspace-based classifiers winners in the context of joint MC and detection. Finally, not shown in this letter for lack of space, are the FER plots with uncorrelated channels, where all schemes achieved near-MT-aware performance, and the CCR plots with imperfect CSI, that show remarkable performance deterioration (mitigating channel estimation errors is beyond the scope of this work).

\section{Conclusion}
\label{sec:conclusion}
Low-complexity per-layer MC schemes have been proposed for MIMO systems, based on subspace decomposition. It has been shown that assuming the modulation type on all layers except the layer of interest to be a dense constellation results in good classification performance. This assumption has been proved to have a negligible performance degradation cost in subspace data detection, which made fully parallelizable efficient joint MC and data detection feasible.

\ifCLASSOPTIONcaptionsoff
  \newpage
\fi

\end{document}